\documentclass[aps,prc,twocolumn,amsmath,amssymb,superscriptaddress,nofootinbib,showpacs,showkeys]{revtex4}
\usepackage{dcolumn}
\usepackage{graphicx}
\usepackage{color}
\usepackage{multirow}
\usepackage{textcomp}
\usepackage{physics}
\usepackage{dsfont}
\usepackage{epstopdf}

\usepackage{relsize}
\begin{document}
	\newcommand {\nc} {\newcommand}
	\nc {\beq} {\begin{eqnarray}}
	\nc {\eeq} {\nonumber \end{eqnarray}}
	\nc {\eeqn}[1] {\label {#1} \end{eqnarray}}
\nc {\eol} {\nonumber \\}
\nc {\eoln}[1] {\label {#1} \\}
\nc {\ve} [1] {\mbox{\boldmath $#1$}}
\nc {\ves} [1] {\mbox{\boldmath ${\scriptstyle #1}$}}
\nc {\mrm} [1] {\mathrm{#1}}
\nc {\half} {\mbox{$\frac{1}{2}$}}
\nc {\thal} {\mbox{$\frac{3}{2}$}}
\nc {\fial} {\mbox{$\frac{5}{2}$}}
\nc {\la} {\mbox{$\langle$}}
\nc {\ra} {\mbox{$\rangle$}}
\nc {\etal} {\emph{et al.}}
\nc {\eq} [1] {(\ref{#1})}
\nc {\Eq} [1] {Eq.~(\ref{#1})}
\nc {\Refc} [2] {Refs.~\cite[#1]{#2}}
\nc {\Sec} [1] {Sec.~\ref{#1}}
\nc {\chap} [1] {Chapter~\ref{#1}}
\nc {\anx} [1] {Appendix~\ref{#1}}
\nc {\tbl} [1] {Table~\ref{#1}}
\nc {\Fig} [1] {Fig.~\ref{#1}}
\nc {\ex} [1] {$^{#1}$}
\nc {\Sch} {Schr\"odinger }
\nc {\flim} [2] {\mathop{\longrightarrow}\limits_{{#1}\rightarrow{#2}}}
\nc {\textdegr}{$^{\circ}$}
\nc {\inred} [1]{\textcolor{red}{#1}}
\nc {\inblue} [1]{\textcolor{blue}{#1}}
\nc {\IR} [1]{\textcolor{red}{#1}}
\nc {\IB} [1]{\textcolor{blue}{#1}}
\nc{\pderiv}[2]{\cfrac{\partial #1}{\partial #2}}
\nc{\deriv}[2]{\cfrac{d#1}{d#2}}
\title{Halo effective field theory analysis of one-neutron knockout reactions of  $^{11}\rm Be$ and $^{15}\rm C$}
\author{C.~Hebborn}
\email{hebborn@frib.msu.edu}
\affiliation{Facility for Rare Isotope Beams, Michigan State University, East Lansing, MI 48824, USA}
\affiliation{Lawrence Livermore National Laboratory, P.O. Box 808, L-414, 
Livermore, California 94551, USA}
\affiliation{Physique Nucl\'eaire et Physique Quantique (CP 229), Universit\'e libre de Bruxelles (ULB), B-1050 Brussels}
\author{P.~Capel}
\email{pcapel@uni-mainz.de}
\affiliation{Institut f\"ur Kernphysik, Johannes Gutenberg-Universit\"at Mainz, D-55099 Mainz}
\affiliation{Physique Nucl\'eaire et Physique Quantique (CP 229), Universit\'e libre de Bruxelles (ULB), B-1050 Brussels}
\date{\today}
\begin{abstract}
\begin{description}
\item[Background] One-nucleon knockout reactions provide insightful information on the single-particle structure of nuclei.
When applied to one-neutron halo nuclei, they are purely peripheral, suggesting that they could be properly modeled by describing the projectile within a Halo Effective Field Theory (Halo-EFT).
\item[Purpose] We reanalyze the one-neutron  knockout measurements of $^{11}$Be and $^{15}$C---both one-neutron halo nuclei---on beryllium at about 60~MeV/nucleon. We consider Halo-EFT descriptions of these nuclei which already provide excellent agreement with breakup and transfer data.
\item[Method] We include a Halo-EFT description of the projectile within an eikonal-based model of the reaction and compare its outcome to existing data.
\item[Results] Excellent agreement with experiment is found for both nuclei.
The asymptotic normalization coefficients inferred from this comparison confirm predictions from \emph{ab initio} nuclear-structure calculations and values deduced from transfer data.
\item[Conclusions] Halo-EFT can be reliably used to analyze one-neutron knockout reactions  measured for halo nuclei and test predictions from state-of-the-art nuclear structure models on these experimental data.
\end{description}
\end{abstract}

\maketitle
%

\section{Introduction}\label{Introduction}
Halo nuclei exhibit an unusually large matter radius compared to their isobars.
These very neutron-rich nuclei are found far from stability close to or at the neutron dripline.
There, the binding of one or two valence nucleons is so loose that they can tunnel far into the classically forbidden region to form an extended diffuse halo around the compact core of the nucleus \cite{HJ87}.
Example of halo nuclei are $^{11}$Be and $^{15}$C seen as a one-neutron halo bound to a $^{10}$Be or $^{14}$C core, and $^6$He and $^{11}$Li with two neutrons in their halo.
Because of their peculiar structure, halo nuclei have been the subject of 
many studies since their discovery in the mid-1980s \cite{Tetal85a,T96}.

Due to their short lifetime, halo nuclei are often studied through reactions such as breakup \cite{Petal03,DPetal03,Fetal04,Netal09}, transfer~\cite{Getal75,Metal11,Setal12,Setal13} and knockout~\cite{Sau00,Aetal00,Tetal02,HT03}.
In the first reaction the halo dissociates from the core during its interaction with the target.
The breakup hence reveals the internal structure of the nucleus.
Thanks to the fragile nature of the projectile, the cross sections are high.
In transfer reactions, the halo state is populated through a $(d,p)$ transfer reaction often measured in inverse kinematics.
Knockout, which is the focus of the present study, corresponds to the removal of the halo neutron from the nucleus.
This reaction therefore contains both the aforementioned breakup---often called \emph{diffractive breakup}---and the absorption of the valence neutron by the target, known as \emph{stripping}.
Because it does not require the detection of a neutron in coincidence with the core, knockout measurement exhibits higher statistics than breakup.
For that reason it is favored over the latter at low beam intensity \cite{HT03,Aetal21}.

The typical observable measured in knockout reactions is the momentum distribution of the core after the collision \cite{Aetal00,Tetal02,HT03,Sau00}.
When measured at high enough energy, the reaction is sudden and that observable retains the memory of the momentum distribution the core had within the nucleus.
Because of the large spatial extension of the core-halo wave function and 
following Heisenberg's uncertainty principle, that distribution is narrow, hence providing a stringent probe of the halo structure \cite{Aetal00,Tetal02,HT03,Sau00}.

Because of their high statistics, the use of knockout reactions has been extended to study the single-particle structure of more deeply-bound exotic nuclei \cite{HT03,Aetal21}. 
Experimental data are usually compared to eikonal calculations of the reaction which include shell-model predictions for the spectroscopic factor of the projectile overlap wave function.
In such analyses, the experimental cross section is usually smaller than what theory predicts \cite{HT03,Aetal21,Gad08,TG14,TG21}.
The quenching factor $R_S$, i.e., the ratio between experiment and theory, is therefore interpreted as quantifying the deviations from shell-model predictions caused by missing correlations in the truncated model space used in structure calculations.
Surprisingly, $R_S$ decreases with the binding energy of the knocked-out nucleon:  $R_S\sim 1$ for the removal of weakly-bound nucleons, as in halo nuclei, and  $R_S\sim0.3$ in the knockout of strongly-bound nucleons \cite{Gad08,TG14,TG21}.
Since this feature is not observed in the analysis of transfer or quasi-free scattering reactions~\cite{Letal10,Aetal18,GRM18,Aetal21}, it remains unclear if it is due to a bias in the interpretation of the knockout data or more profound differences between these reaction mechanisms and how 
they should be analyzed and compared to one another.

In a previous analysis, we have studied the sensitivity of the knockout cross sections of one-neutron halo nuclei to the description of their structure  \cite{HC19}. 
We have shown that these reactions are peripheral, in the sense that they 
probe only the tail of the core-halo wave function.
Moreover, they are quite insensitive to the description of the core-neutron continuum.
Accordingly, to properly reproduce experimental data, the model of the reaction should include a description of the projectile that reproduces its 
binding energy, the correct asymptotic normalization coefficient (ANC), and excited bound states and their ANC.
For this goal, the halo effective field theory (Halo-EFT~\cite{BHvK02,BHvk03}, see Ref.~\cite{HJP17} for a recent review) at next-to-leading order 
(NLO) includes all the relevant structure information.
Interestingly, there already exist such Halo-EFT descriptions for $^{11}$Be \cite{CPH18} and $^{15}$C \cite{MYC19}.
They are constrained with experimental one-neutron separation energies and the ANCs of the bound states predicted by \textit{ab initio} calculations \cite{Cetal16,Navratil18} or inferred from analyses of transfer reactions \cite{YC18,MYC19}.
These descriptions successfully reproduce both breakup \cite{CPH18,MC19,MYC19} and transfer \cite{YC18,MYC19} data.
In this work, we use these Halo-EFT descriptions to reanalyze knockout parallel-momentum distributions and integrated cross sections for $^{11}$Be 
and $^{15}$C on a $^9$Be target at, respectively, 60~MeV/nucleon and 54~MeV/nucleon \cite{Aetal00,Tetal02}. 

We briefly present in Sec.~\ref{Sec2} the reaction model, including the Halo-EFT descriptions of  $^{11}$Be and $^{15}$C and our choices of optical potentials.
In Sec.~\ref{Sec3}, we compare our theoretical predictions with experimental data and investigate the uncertainty associated with the choice of optical potentials.
We conclude in Sec.~\ref{Conclusions}.

\section{Few-body model of knockout}\label{Sec2}

\begin{table*}
	\centering
	\begin{tabular}{cccccccccccccc} \hline \hline
		\multicolumn{2}{c}{Interaction} &$E_{\rm beam}$&$V_R$  & $r_R$& $a_R$ & 
$W_I$ & $r_I$ &$a_I$  &$W_D$ & $r_D$  &$a_D$ &$r_C$ & Ref. \\
		& &[MeV]&[MeV]&[fm]&[fm]&[MeV] & [fm]&[fm]&[MeV]& [fm]&[fm]&[fm]&\\ \hline
		\multirow{2}{*}{$c$-$^{9}\mathrm{Be}$}&$V_{cT}^1$&&123.0& 0.75& 0.80&  65.0& 0.78&  0.80&0 &0&0&1.2 &\cite{ATB97}\\
		&$V_{cT}^2$&& 127.0& 0.80&0.78&13.90&  1.25& 0.70 &0 &0&0&1.0&\cite{Lietal13}\\\hline
		\multirow{3}{*}{$n$-$^{9}\mathrm{Be}$}&$V_{nT}^1$&&33.08 & 1.14& 0.65&4.15& 1.14 & 0.65 &9.18 &2.83  &0.18& &\cite{W18}\\
		&\multirow{2}{*}{$V_{nT}^2$}&60&22.60 & 1.37& 0.29& 5.30 & 1.30&0.30 &15.25 &1.30 & 0.30& &\multirow{2}{*}{\cite{BC14}}\\
		&&54&23.47&  1.40&0.29& 5.36 & 1.30& 0.30  &15.55&1.30 & 0.30 & &\\ \hline 
\hline
	\end{tabular}	
	\caption{Parameters of the optical potentials \eq{eq6} used to simulate the $c$-$^9\rm Be$ and $n$-$^9\rm Be$ interactions for the one-neutron knockout of $^{11}\rm Be$ and $^{15}\rm C$ on $^9\rm Be$ at 60~MeV/nucleon and 54~MeV/nucleon.
To estimate the uncertainty related to this choice, two potentials have been selected for each interaction.}
	\label{t1}
\end{table*}

\subsection{Reaction model}

To model the one-neutron knockout of halo nuclei, we rely on the usual few-body framework \cite{HT03,BC12}.
The projectile is described as a two-body system: a core $c$, assumed to be in its $0^+$ ground state, to which the halo neutron $n$ is loosely bound.
The structure of the target $T$ is neglected.
The $c$-$n$ interaction is modeled by an effective single-particle potential whose parameters are fitted to reproduce the lower-energy spectrum of 
the projectile: the energy, spin and parity of its bound states and of some of its resonances above the $c$-$n$ separation threshold.
The interaction between the projectile constituents and the target are simulated by optical potentials, which include an imaginary term that accounts for all the inelastic channels not explicitly accounted for by the model.

Within this three-body framework, we evaluate the knockout cross sections 
using eikonal-based models of the collision \cite{G59}.
Two processes contribute to the knockout observables: the diffractive breakup, in which both the core and the halo neutron survive the collision, and the stripping, in which the neutron is absorbed by the target.
The diffractive-breakup contribution is computed within  the dynamical eikonal approximation (DEA) \cite{BCG05,GBC06}, which provides a proper dynamical description of the collision.
Because the DEA has not yet been extended to stripping observables, we compute these contributions at the usual eikonal approximation~\cite{HM85,HT03,HBE96}. Contrary to the DEA, the usual eikonal model relies on the adiabatic approximation, which sees the internal coordinates of the projectile as frozen during the collision.
This approximation neglects the dynamical effects in the stripping.

All calculations are obtained with the same model spaces as the ones detailed in Refs.~\cite{CBS08,HC19}.  
\subsection{Halo-EFT description of $^{11}\rm Be$ and $^{15}\rm C$}

We describe $^{11}$Be and $^{15}$C within Halo-EFT.
This approach exploits the separation of scales between the size of the halo and that of the core to expand the projectile Hamiltonian upon the small parameter $R_{\rm core}/R_{\rm halo}$.
In this expansion, both the core and halo neutron are considered structureless and the breakdown scale is set by the size of the core \cite{BHvK02,BHvk03,HJP17}.
The short-range physics is absorbed by a contact interaction and its derivatives, whose parameters are adjusted in each partial wave to reproduce known long-range properties of the nucleus, such as its one-neutron separation energy and the ANC of its bound states.
For practical handling within reaction codes, these interactions are regulated with a Gaussian (see Eq.~(13) of Ref.~\cite{CPH18}).

Both $^{11}$Be and $^{15}$C  exhibit a similar structure.
They both have a $1/2^+$ ground state, which is seen as a $1s_{1/2}$ neutron bound to their $^{10}\rm Be$ or $^{14}\rm C$ core in its $0^+$ ground state.
Their spectra also include a subthreshold excited state: a $1/2^-$ state 
for $^{11}\rm Be$ and a $5/2^+$ state for $^{15}\rm C$.
These states are also described as a neutron bound to the $0^+$ core, in either the $0p_{1/2}$ or the $0d_{5/2}$ orbital, respectively.
Following Refs.~\cite{CPH18,MYC19}, we describe $^{11}\rm Be$ and $^{15}\rm C$ at next-to-leading order (NLO).
For $^{11}\rm Be$, we set an effective $c$-$n$ potential in the $s_{1/2}$ 
and $p_{1/2}$ waves, which reproduces the experimental one-neutron separation energies in the $1/2^+$ and  $1/2^-$ states, as well as the ANCs predicted by the \emph{ab initio} calculations of Calci {\etal} \cite{Cetal16}.
In the present work, we consider the potential with  the Gaussian range of $1.2$~fm, for which the  depths can be found in Tables~I and~II of Ref.~\cite{CPH18}.
The uncertainty associated with the choice of this range is negligible, i.e., less than 1.5\% for integrated knockout cross sections.
We do not include any interaction in the $p_{3/2}$ or higher partial waves, first to reproduce the nearly nil phaseshifts predicted \emph{ab initio} at low $^{10}$Be-$n$ energy \cite{Cetal16}, and second because the details of the core-neutron interaction in the continuum do not matter in such calculations \cite{HC19}.

In the case of $^{15}\rm C$, we use an effective potential in the $s_{1/2}$ partial wave to reproduce the ground state and go beyond NLO by including an interaction within the $d_{5/2}$ wave to describe the bound excited state, that can affect knockout calculations \cite{HC19}.
 In this article, we use the parameters of these interactions for a Gaussian range of 1.2 fm,  which are given in Table~III of Ref.~\cite{MYC19}. As for $^{11}\rm Be$ case, the uncertainty associated with this range are negligible.
These potentials reproduce the ANCs inferred from an analysis of transfer 
observables \cite{MYC19}, and agree very well with those predicted by an \emph{ab initio} calculation~\cite{Navratil18}.
As in Ref.~\cite{MYC19}, we assume a nil $c$-$n$ interaction in the other 
partial waves, in particular the $p$ ones.
First we do not have any reliable prediction for the corresponding phaseshifts and second we know from our previous study that the knockout cross sections are insensitive to the particulars of the $c$-$n$ continuum \cite{HC19}.

Once coupled with accurate reaction models, these Halo-EFT descriptions of $^{11}$Be and $^{15}$C lead to an excellent agreement with breakup and transfer data \cite{CPH18,YC18,MC19,MYC19}.
The goal of the present study is to see if, following our previous study \cite{HC19}, they can also explain the knockout cross sections measured in Refs.~\cite{Aetal00,Tetal02}.

\subsection{Optical potentials}\label{Sec2b}

To analyze one-neutron knockout reactions of $^{11}$Be and $^{15}$C on $^9$Be at 60~MeV/nucleon and 54~MeV/nucleon, respectively \cite{Aetal00,Tetal02}, we need optical potentials to simulate the interaction between the 
projectile constituents and the target.
To estimate the uncertainty related to that choice, we select in the literature two different optical potentials for each interaction.

The first $c$-$^9\rm Be$  potential $V_{cT}^1$, developed  in Ref.~\cite{ATB97}, reproduces elastic-scattering data of $^{10}\rm Be$ on $^{12}\rm C$ at a beam energy $E_{\rm beam}=59.4$~MeV/nucleon~\cite{CGPhD}.
Although it does not correspond to the exact same target, it provides a fair estimate for the $^{10}$Be-$^9$Be interaction at the right beam energy.
Not having found a more adequate potential for the $^{14}$C-$^9$Be interaction, we use this potential also in that latter case.
The second $c$-$^9\rm Be$ potential $V_{cT}^2$ has been adjusted on elastic-scattering data for $^9\rm Be$ on $^{13}\rm C$ at 40~MeV \cite{Lietal13}.
Although it has been fitted at a beam energy much lower than those considered here, it is the second most realistic potential we could find in the 
literature.

For the first $n$-$^9\rm Be$ interaction $V_{nT}^1$, we adopt the global optical potential developed by Weppner \cite{W18}.
It has been fitted to elastic-scattering angular distributions and polarization data for a nucleon off a nucleus with mass number $A\leq13$ at energies between 65~MeV and 75~MeV.
This potential is therefore well adapted for the collisions studied here.
The second  $n$-$^9\rm Be$ potential $V_{nT}^2$, parametrized by Bonaccorso and Charity, reproduces the total, elastic, and reaction cross sections for a neutron on $^9\rm Be$ at energies between 1 and 100 MeV \cite{BC14}.
We compute its parameters for the two beam energies considered here.

\begin{figure*}
	\centering
	{\includegraphics[width=0.48\linewidth]{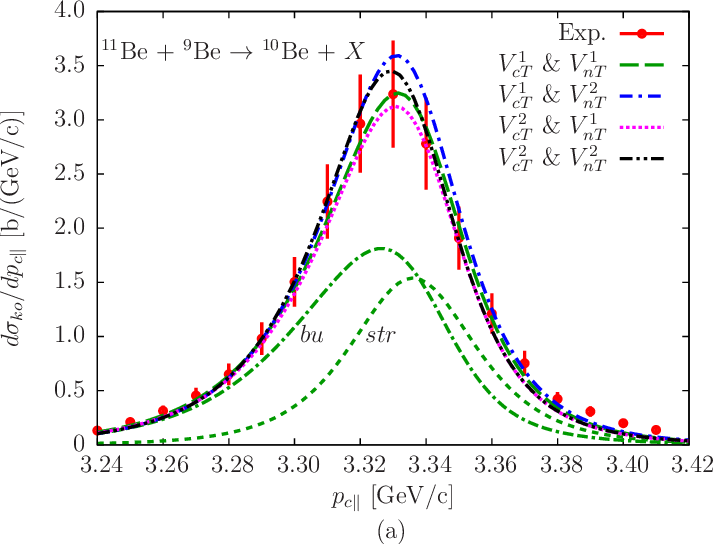}}\hspace{0.4cm}
	{\includegraphics[width=0.48\linewidth]{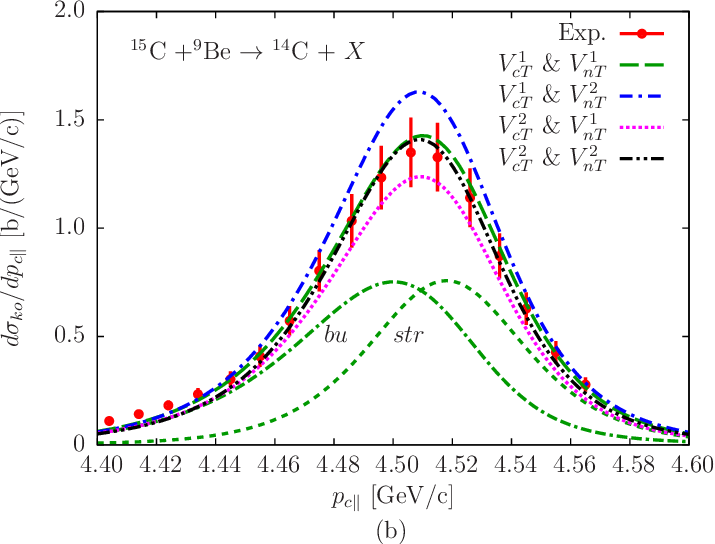}}
	\caption{{Parallel-momentum distributions for the one-neutron knockout on 
$^9\rm Be$ of (a) $^{11}\rm Be$ at 60~MeV/nucleon, and (b) $^{15}\rm C$ at 54~MeV/nucleon. Calculations obtained with the different optical potentials of Table~\ref{t1} are compared to the experimental data of Refs.~\cite{Aetal00,Tetal02}. The experimental error bars correspond to the 15\% and 12\% uncertainty cited in Refs.~\cite{Aetal00,Tetal02}, respectively.
For the first choice of optical potentials, the diffractive-breakup ($bu$, double-dash-dotted green line) and stripping ($str$, short-dashed green line) contributions to the cross section are shown separately.
}}\label{f1}
\end{figure*}

The nuclear part of all these potentials exhibit the usual expression
\begin{eqnarray}
V (R) &=& -V_R\,f_{\rm WS}(R,R_R,a_R) -i\,W_I\,f_{\rm WS}(R,R_I,a_I) \nonumber \\
&& +i\,4 a_D W_D \deriv{}{R} f_{\rm WS}(R,R_D,a_D), \label{eq6}
\end{eqnarray}
with $f_{\rm WS}(R,R_X,a_X) = \frac{1}{1 +e^{\frac{R-R_X}{a_X}}}$ the Woods-Saxon form.
For the $c$-$T$ interaction, the radii are parametrized as $R_x=r_x(A_c^{1/3}+A_T^{1/3})$, with $A_c$ and $A_T$ the core and  target mass numbers, respectively.
The radii for the $n$-$T$ interactions are obtained with $R_x=r_x\times 
A_T^{1/3}$. 
The Coulomb $c$-$T$ interaction is simulated by the potential generated
by a uniformly charged sphere of radius $R_C=r_C\times (A_c^{1/3}+A_T^{1/3})$.
The parameters of the optical potentials considered in this study are listed in Table~\ref{t1}.

\section{Analysis of the one-neutron knockout from $^{11}$Be and $^{15}$C}\label{Sec3}

To see if Halo-EFT descriptions of $^{11}$Be and $^{15}$C at NLO correctly reproduce one-neutron knockout, we perform reaction calculations within 
the model described in \Sec{Sec2} and compare our results with the data of Refs.~\cite{Aetal00,Tetal02} obtained on a $^9$Be target at 60~MeV/nucleon and 54~MeV/nucleon, respectively.
To estimate the uncertainty related to the simulation of the interaction between the projectile constituents and the target, we consider the four possible combinations of the optical potentials listed in Table~\ref{t1}.
For each of these choices, the total one-neutron knockout cross section {($\sigma_{ko}$)} is provided in Table~\ref{t2},
{as well as its two contributions: diffractive breakup ($\sigma_{bu}$) and stripping ($\sigma_{str}$).}
{The experimental values of Refs.~\cite{Aetal00,Tetal02} are also listed.}
{The knockout cross sections expressed as a function of the momentum of the core parallel to the beam axis $p_{c\parallel}$ are displayed in \Fig{f1}.
For the first combination of optical potentials, the diffractive-breakup and stripping contributions are shown separately.}
The experimental data (red points) are extracted from Figs.~11 and 12 of  
Ref.~\cite{Tetal02}.
The error bars correspond to the 15\% and 12\% experimental uncertainties reported in Refs.~\cite{Aetal00,Tetal02}, respectively.
For a proper comparison between theory and experiment, following Ref.~\cite{Tetal02}, we have adjusted the  position of the center of each parallel-momentum distributions to the data.
Beside that minor adjustment, no parameter has been fitted to the data; in particular, the magnitude of the cross sections is the direct output of 
the calculations.

\begin{table*}
	\centering
	\begin{tabular}{cc|c|cccc} \hline \hline
	&	&Exp.&$V_{cT}^1$ \& $V_{nT}^1$ &$V_{cT}^1$ \& $V_{nT}^2$ &$V_{cT}^2$ \& 
$V_{nT}^1$ & $V_{cT}^2$ \& $V_{nT}^2$ \\ \hline
		\multirow{3}{*}{$^{11}\rm Be$}&$\sigma_{bu}$&&113&89&110&89\\
		&$\sigma_{str}$&&81&118&74&108\\
		&$\sigma_{ko}$& 203$\pm$31&194&207&185&197\\  \hline
		\multirow{3}{*}{		$^{15}\rm C$}&$\sigma_{bu}$&&60&48&52&43\\
		&$\sigma_{str}$&&51&75&43&63\\
		&$\sigma_{ko}$&109$\pm$13&111&124&95&105\\ \hline\hline
	\end{tabular}	
	\caption{{Integrated diffractive-breakup ($\sigma_{bu}$), stripping ($\sigma_{str}$)  and knockout ($\sigma_{ko}=\sigma_{bu}+\sigma_{str}$)} cross sections in mb for $^{11}\rm Be$ and  
$^{15}\rm C$ on $^9\rm Be$ at 60~MeV/nucleon and 54~MeV/nucleon, respectively. Theoretical results obtained with the different potentials listed in Table~\ref{t1} are compared to the experimental values from Refs.~\cite{Aetal00,Tetal02}.}
	\label{t2}
\end{table*}

We observe an excellent agreement between theory and experiment for both knockout observables.
This is first seen in the magnitude of the cross section.
In both Table~\ref{t2} and Fig.~\ref{f1}, but for two exceptions, all calculations fall within one standard deviation of the experimental value.
In addition, our calculations reproduce very well the experimental parallel-momentum distribution. 
As deduced from the initial studies, this confirms the clear one-neutron halo structure of $^{11}$Be and $^{15}$C \cite{Aetal00,Tetal02,HT03}.

Because our results are nearly insensitive to the details of the Halo-EFT description of the projectile, the dominant uncertainty in the reaction model resides in the choice of the optical potentials.
This choice affects mostly the magnitude of the cross section, leading to 
about 11\% difference in the $^{11}$Be case and up to 26\% in the $^{15}$C calculations,
{which is much larger than the 1.5\% uncertainty on the $c$-$n$ interaction we have observed.}
For the former nucleus, most of that uncertainty comes from $V_{nT}$, whereas for the latter both $c$-$T$ and $n$-$T$ interactions share an equal role in the changes in the cross section.
Reducing that uncertainty requires the development of new, more accurate, optical potentials, e.g., derived from first principles \cite{Rot18,DCH18,WLH19,WLH20}.
A systematic study within a Bayesian approach, such as done in Refs.~\cite{LNSW17,LN18,Ketal19} for transfer reactions, would help us better understand the influence of each interaction on the reaction process and provide a more reliable uncertainty.
Although in a much less significant way, the optical potentials also influence the shape of the parallel-momentum distribution.
The $n$-$^9$Be optical potential of Bonaccorso and Charity \cite{BC14} produces a slightly less asymmetric peak than Weppner's \cite{W18}.
Albeit not statistically significant, this hints at the influence of the $n$-$T$ interaction in those reactions dynamics \cite{HC21}.

{To study further the role of the optical potentials, we separate the knockout cross section into its diffractive-breakup ($bu$) and stripping ($str$) components in Table~\ref{t2}.
As in Ref.~\cite{Tetal02}, we note that each component contributes for roughly half of the knockout cross section.
The actual division into diffractive breakup and stripping depends mostly on $V_{nT}$: the less absorptive $V_{nT}^1$ of Ref.~\cite{W18} leads to a larger $\sigma_{bu}$ and smaller $\sigma_{str}$ than the more absorptive $V_{nT}^2$ of Ref.~\cite{BC14}.
The $c$-$T$ interaction plays only a minor role in this division.}

{The separation of the $p_{c\parallel}$ distribution into its two contributions is illustrated in Fig.~\ref{f1} for the first choice of optical potentials (the other choices lead to similar results, but for the relative magnitude of the two contributions, see Table~\ref{t2}).
The diffractive-breakup contribution is plotted in double-dash-dotted green line and the stripping one in short-dashed green line.
The former exhibits a notably asymmetric shape with a tail extending to the low-momentum side.
This asymmetry is due to the dynamics of the reaction \cite{Tetal02,CBS08}, which is correctly accounted for within the DEA \cite{BCG05,GBC06}. 
The stripping cross section, on the contrary is purely symmetric around its maximum because it is computed within the usual eikonal approximation, i.e., including the adiabatic approximation \cite{Tetal02}.
Although their sum reproduces very well the experimental cross section, our calculations slightly underestimate the data in the low-momentum tail, especially for the knockout of $^{15}$C.
Our description of the reaction therefore misses some dynamical effects.
To account for the reaction dynamics in the stripping process, the authors of Ref.~\cite{Tetal02} assume the same shape of that parallel-momentum distribution as the diffractive-breakup one.
Their methodology hence includes more dynamical effects than ours; accordingly the difference between their calculations and the data at low momenta is smaller than ours.
A more precise treatment of the dynamics within the stripping reaction is thus needed to properly describe the full reaction process.
Nevertheless, but for this minor issue, our treatment of the reaction mechanism is accurate enough to analyze these measurements.}

Since knockout is purely peripheral for halo nuclei, the cross section scales very well with the square of the ANC of the ground state \cite{HC19}.
From our calculations, we can infer an ANC by adjusting the magnitude of our parallel-momentum distributions to the data.
For $^{11}\rm Be$ we obtain an ANC$^2=0.62 \pm 0.06\pm 0.09$~fm$^{-1}$
and for $^{15}\rm C$, an ANC$^2=1.57 \pm0.30\pm0.18$~fm$^{-1}$.
The first quoted uncertainty is associated with the choice of optical potentials, while the second is due to the experimental error.
Because they are of the same order of magnitude, reducing the uncertainty 
on the ANC inferred from knockout data would require better constraints on the optical potentials and more precise data.

These ANCs inferred from the knockout data are in excellent agreement with the ones predicted by \emph{ab initio} calculations, viz. 0.618~fm$^{-1}$ for $^{11}\rm Be$ \cite{Cetal16} and 1.644~fm$^{-1}$ for $^{15}\rm C$ \cite{Navratil18}, and with Halo-EFT analyses of transfer reactions, viz. $0.616\pm0.001$~fm$^{-1}$ for $^{11}\rm Be$ \cite{YC18} and  $1.59\pm0.06$~fm$^{-1}$ for $^{15}\rm C$ \cite{MYC19}.
Although in the lower end of the spectrum, the ANC for $^{15}\rm C$ agrees well with the values extracted from other analyses~\cite{Metal11,SN08,SN08Err,Tetal06,McCetal14,PNM07,TetalTAM02} (see list in  Table II of Ref.~\cite{MYC19}).
The values listed above have been used to constrain the Halo-EFT $c$-$n$ interactions considered in this analysis as well as in Refs.~\cite{CPH18,MC19,MYC19}:
they lead to an excellent agreement with data for breakup at both intermediate \cite{CPH18,MYC19} and high~\cite{MC19,MYC19} energies, transfer \cite{YC18,MYC19}, and the radiative capture $^{14}{\rm C}(n,\gamma)$ at energy of astrophysical interest~\cite{MYC19}.
This result is thus the last piece of a puzzle, which demonstrates that with one Halo-EFT description of halo nuclei, we can reproduce various sets of experimental data using accurate reaction models.
Accordingly, Halo-EFT properly identifies the nuclear-structure observables that matter most in these reactions, viz. the binding energy of the halo neutron, its orbital angular momentum and the ANC of the projectile ground state.
In particular, the norm of the overlap wave function does not influence the results of the calculations, as long as the ANC is properly reproduced 
\cite{CPH18,HC19}.

\section{Conclusions}\label{Conclusions}

Knockout reactions provide a clean probe of halo nuclei because their cross section reveals clearly the spatial extension of the nuclear wave function and they exhibit high statistics \cite{Aetal00,Tetal02,HT03,Sau00}.
Our previous analysis~\cite{HC19} has shown that the one-neutron knockout 
of halo nuclei is  sensitive only to the asymptotics of the ground-state wave function while being insensitive to the norm of that wave function, i.e., the spectroscopic factor of the corresponding overlap wave function.
Moreover, knockout cross sections do not depend on the description of the 
$c$-$n$ continuum.
Accordingly, a Halo-EFT description at NLO of the projectile should include all the relevant nuclear-structure information to reproduce such data.
In this article, we test this idea by reanalyzing measurements of one-neutron knockout of $^{11}\rm Be$ \cite{Aetal00} and $^{15}\rm C$ \cite{Tetal02} using Halo-EFT descriptions reproducing well data on breakup \cite{MYC19,CPH18,MC19} and transfer~\cite{YC18,MYC19}.

For both nuclei, our calculated parallel-momentum distributions and integrated knockout cross sections are in excellent agreement with the data.
In particular, the typical narrow width of the distributions, their slight asymmetry, and the magnitude at the peak are well reproduced.
Using four different optical potentials, we have estimated the uncertainty of the reaction model and suggested that more accurate optical potentials would reduce that uncertainty \cite{Rot18,DCH18,WLH19,WLH20}.

Comparing our calculations with the data, we have inferred the ground-state ANC for both projectiles.
These values agree well with \emph{ab initio} predictions \cite{Cetal16,Navratil18} and transfer analyses \cite{YC18,MYC19}.
Our results complete a series of previous studies and demonstrate that one description of $^{11}\rm Be$ and $^{15}\rm C$ based on Halo-EFT reproduces independent experimental data for breakup~\cite{CPH18,MC19,MYC19}, transfer~\cite{YC18,MYC19}, radiative-capture \cite{MYC19} and, from this study, knockout \cite{Aetal00,Tetal02} measurements. 
Since our results are independent of the normalization of the ground-state wave function~\cite{HC19}, they also suggest that the quenching factor 
$R_S\sim1$ obtained for halo nuclei in previous analyses of knockout data 
\cite{Gad08,TG14,TG21} is due more to the use of realistic ANCs in reaction calculations than to accurate spectroscopic factors.
This is not surprising since, for neutrons loosely bound in an $s$ wave, the ANC of the wave function is predominantly determined by the binding energy, with little sensitivity to the  $c$-$n$ potential geometry \cite{SCB10}.

\begin{acknowledgements}
		The work of C.~Hebborn was supported by the Fund for Research Training in 
	Industry and Agriculture (FRIA) and  by the U.S. Department of Energy, Office of Science, Office of Nuclear Physics, under the FRIB Theory Alliance award no. DE-SC0013617 and {under Work Proposal no. SCW0498}. This work was prepared in part by LLNL under Contract no. DE-AC52-07NA27344.    This project has received funding from the European Union’s Horizon 2020 research	and innovation program under grant agreement 	No 654002,    the Fonds de la Recherche Scientifique - FNRS under Grant Number 4.45.10.08,  the Deutsche Forschungsgemeinschaft Projekt-ID 279384907 – SFB 1245 and Projekt-ID 204404729 – SFB 1044, and the PRISMA+ (Precision Physics, Fundamental Interactions and Structure of Matter) Cluster of Excellence. P. C. acknowledges the support of the State of Rhineland-Palatinate.
\end{acknowledgements}

\end{document}